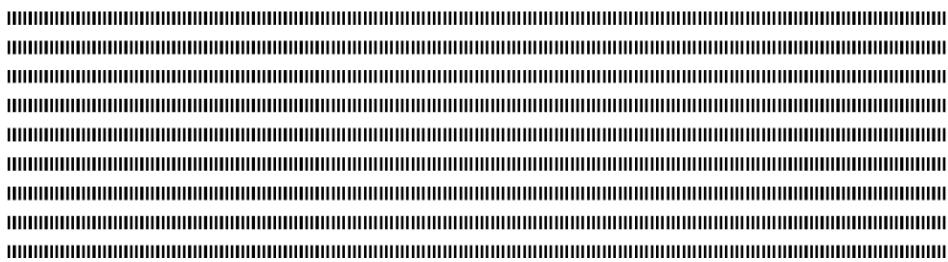

# Analyse formelle de concept pour le routage des requêtes dans les systèmes pair-à-pair

Taoufik Yeferny, Khedija Arour

Département des sciences informatiques
Faculté des Sciences de Tunis
Tunis Tunisie

Institut National des Sciences Appliquées et de Technologie de Tunis
Tunis
Tunisie

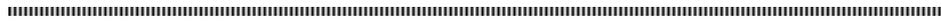

**RÉSUMÉ.** Les systèmes pair-à-pair (P2P) ont remporté ces dernières années un succès rapide auprès du grand public et des professionnels car ils offrent la possibilité aux utilisateurs de partager et d'accéder à des ressources diverses, distribuées à large échelle. Actuellement, les systèmes P2P constituent des grilles d'informations regroupant des millions des pairs connectés. Par conséquent, la sélection des pairs qui peuvent répondre aux requêtes des utilisateurs devient un problème très important. De nombreuses recherches concernant la sélection des meilleurs pairs contenant les données appropriées à une requête, ont émergé et constituent un axe de recherche très actif. Nous présentons dans ce papier une amélioration de notre algorithme de routage des requêtes par apprentissage LearningPeerSelection (LPS) présenté dans CORIA 2009, une nouvelle méthode de construction des bases de connaissances et étude expérimentale avancée. Pour tester l'algorithme proposé, nous avons défini une couche de routage sur le simulateur PeerSim qui nous a permis d'évaluer l'efficacité de notre algorithme.

**ABSTRACT.** The Peer-to-Peer systems (P2P) were led these last years as the major technology of access upon various resources on Internet. These systems build a cluster witch contains a very large number of peers. As the result the selection of peers who can answer for a given query is a very difficult problem. The efficiency of the selection algorithms can be improved by introducing of semantics into the process of queries routing. We present in this paper a novel improved version of our semantic routing algorithm *LearningPeerSelection* (*LPS*) presented in CORIA 2009, an incremental strategy of updating knowledge bases and an advanced experimental study. To test the proposed algorithm, we defined a layer of routing on the PeerSim simulator.

**MOTS-CLÉS :** Systèmes P2P, Routage sémantique, apprentissage, Analyse des Concepts Formels, Simulateur PeerSim.

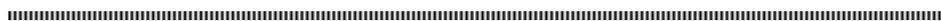

## 1. Introduction

Les systèmes pair-à-pair (P2P) ont remporté ces dernières années un succès rapide auprès du grand public et des professionnels car ils offrent la possibilité aux utilisateurs de partager et d'accéder à des ressources diverses, distribuées à large échelle. Actuellement les systèmes de recherche d'information P2P (SRI P2P), constituent des grilles d'informations regroupant des millions des pairs connectés et tout pair participant peut partager un ensemble de ses documents comme il peut aussi accéder à n'importe quel document ou fichier partagé par les autres pairs. Par conséquent, la sélection des pairs qui peuvent répondre aux requêtes des utilisateurs devient un problème très important. Ce problème est appelé problème de routage des requêtes dans les SRI P2P.

Nous pouvons classifier ces systèmes selon leur modèle de recherchesous-jacent, qui peut être soit non structuré (propagation aléatoire des requêtes dans le graphe des pairs), soit structuré (propagation des requêtes selon une structure d'organisation des pairs basée en général sur des fonctions de hachage). Chaque type de système possède des avantages et des inconvenants.Parmi les avantages des systèmes non structurés, nous pouvons signaler le fait qu'ils respectent au mieux l'autonomie des pairs et ils supportent des langages de requêtes plus expressifs. Cependant, ils ne sont pas efficaces sur le plan du routage des requêtes. Plusieurs travaux de recherche ont tenté d'améliorer la méthode de routage des requêtes dans ces systèmes en introduisant la sémantique dans le processus de propagation de requête [3, 4]. Cette sémantique est généralement construite à partir du contenu des pairs, mais peut également faire intervenir le comportement explicite des utilisateurs. Nous présentons dans ce papier une amélioration de notre algorithme de routage des requêtes par apprentissage LearningPeerSelection (LPS) présenté dans CORIA 2009, une nouvelle méthode de construction des bases de connaissances et étude expérimentale avancée.

Le reste de l'article est organisé comme suit. Dans la section [1], nous présentons un état de l'art sur le routage des requêtes dans les SRI P2P. Les sections 3 et 4 présentent une amélioration l'algorithme *LPS*. Les résultats des différentes expérimentations de l'algorithme seront discutés dans la section 5. Enfin, nous concluons cet article par une conclusion et quelques perspectives intéressantes de prolongement de notre travail.

local. S'il ne peut pas honorer la requête, il calcule les rangs des pairs de son index global et il propage la requête aux pairs de plus grand rang. La terminaison de la recherche est assurée par une détection de cycles ainsi qu'une profondeur maximale de recherche (appelée TTL ou Time To Live).

## [1] . Etat de l'art sur le routage des requêtes dans les systèmes P2P

L'objectif d'une méthode de routage est de router les requêtes automatiquement vers l'ensemble des pairs susceptibles de fournir une réponse. Dans la littérature, plusieurs travaux de recherche ont essayé d'améliorer la méthode classique de routage des requêtes, qui propage une requête à un ensemble de pairs choisit d'une manière aléatoire, en introduisant la sémantique dans le processus de propagation des requêtes [3, 2, 4]. Nous pouvons classifier les méthodes de routage sémantique en méthodes basées sur le contenu et méthodes basées sur l'historique des requêtes. Parmi celles qui sont basées sur le contenu de pairs, nous pouvons citer la méthode PlanetP [9], qui construit au niveau de chaque pair un index local décrivant sa collection locale et un index global contenant des metainformations sur les documents partagés par les autres pairs, pour lui donner une vision résumée des documents partagés par les autres. L'algorithme de routage est très simple, un noeud qui reçoit une requête commenced'abord par faire une recherche dans son index

Bien que cette solution ait donné des meilleurs résultats que la méthode classique, elle ne permet le passage à l'échelle du système. L'espace de stockage nécessaire pour l'index est également trop important et sa mise à jour est difficile, elle implique trop de trafic sur le réseau et une surcharge de travail pour les pairs. Peu de méthodes utilisent l'information sur l'historique des requêtes. Néanmoins, la méthode REMINDIN [3] exploite les métaphores sociales pour définir une stratégie de routage des requêtes basée sur l'historique des requêtes. Pour cela, elle utilise une ontologie (un dépôt local) au niveau de chaque pair. Ces ontologies, sont des déclarations RDF [12] construites à partir des réponses aux requêtes passées. En plus de ces déclarations RDF, des méta-informations concernant celles-là sont stockées afin de mémoriser la source de cette déclaration et la valeur de confiance dans la connaissance d'un pair à une ressource ainsi que la valeur de confiance globale dans le pair. Le choix des pairs est basé sur des observations de la connaissance des autres pairs. En effet, l'évaluation de requête, en utilisant l'ontologie associée à un pair renvoie un ensemble de pairs triés selon les valeurs de confiance. Si le nombre de pairs choisis est insuffisant (inférieur à un seuil), la méthode REMINDIN rafraîchit la requête et répète le même processus plusieurs fois. Les valeurs de confiance sont mises à jour à la réception des réponses. Souvent la réponse à une requête contient plus d'informations que l'on peut maintenir dans le dépôt local, l'utilisateur doit alors déterminer les informations à stocker. Un problème très important se pose pour cette approche qui suppose avoir une ontologie partagée par l'ensemble des pairs, ce qui n'est pas toujours disponible pour la communauté des systèmes P2P. De même, l'autre problème est l'intervention de l'utilisateur pour sélectionner les déclarations qui sont en relation avec la requête afin de construire le dépôt local.

## 3. Algorithme : LearningPeerSelection

### 3.1. Présentation

La méthode de sélection des pairs de l'algorithme *LPS* supporte des nouvelles connaissances que nous avons désignées par profil. Les profils sont construits au niveau de chaque pair d'une manière hors ligne ce qui permet de supporter le facteur d'échelle. Cette solution est basée sur l'architecture du système Gnutella [6] qui est le système le plus représentatif des systèmes de recherche d'information P2P non structurés. Gnutella est un système de partage de fichiers sur Internet qui représente aujourd'hui l'archétype des systèmes P2P non structurés. Dans ce système, une requête est d'abord évaluée sur le pair d'émission puis propagée récursivement sur un sous-ensemble aléatoire des pairs voisins (principe d'inondation). La terminaison de la recherche est assurée par une détection de cycles ainsi qu'une profondeur maximale de recherche (appelée *TTL* ou Time To Live).
Dans le but d'améliorer l'efficacité de cette méthode, le module de propagation des requêtes a été remplacé par un autre module de propagation"guidée" qui utilise une base de connaissances afin de router les requêtes vers les pairs "pertinents". De plus deux autres modules ont été ajoutés :

– Un module de gestion du fichier log : A la réception d'une réponse, ce modulemet à jour le fichier log en ajoutant les informations relatives à cette requête, à savoir l'identifiant de la requête, l'ensemble de ses termes, les documents téléchargés et les pairs associés.

– Un module de gestion de profils qui s'exécute périodiquement. Son rôle consiste àconstruire au niveau de chaque pair une base de connaissances à partir des informations extraites de l'historique des requêtes. Les profils représentent des corrélations sémantiques entre les requêtes passées et les pairs positifs (i.e les pairs à partir desquels y a eu des téléchargements). En effet, le but est de trouver des groupements des requêtes qui partagent les mêmes termes et des groupements des pairs qui ont agi positivement à un ensemble de requêtes. Pour ce faire, il a eu recours à une approche formelle, l'Analyse Formelle de Concepts (FCA) [5], qui répond bien a ce problème. Deux contextes formels sont utilisés, le premier représente le lien entre les requêtes passées et les termes associés, appelé contexte $C1$. Le second, représente le lien entre les requêtes passées et les pairs positifs, appelé contexte $C2$. Un algorithme de génération de concepts formels est par la suite appliqué pour générer deux ensembles de concepts, notés par $E1$ et $E2$. Les concepts de l'ensemble $E_1$, respectivement de $E_2$, seront sous la forme suivante $(\{R_1,..,R_i\},\{T_1,..,T_k\})$, respectivement, $(\{R_1,..,R_i\},\{P_1,..,P_j\})$ avec $\{R_1,..,R_i\}$ un ensemble d'identifiants de requêtes, $\{T_1,..,T_k\}$ un ensemble de termes et $\{P_1,..,P_j\}$ un ensemble de pairs. Ces deux ensembles constituent une base $B(E_1,E_2)$ qui servira par la suite comme une base de connaissances pour l'algorithme de sélection des pairs.

Pour prendre en compte les nouvelles informations sur les requêtes passées, les bases des connaissances sont mises à jour après l'émission d'un certain nombre de requêtes. Pour ce faire, la base de connaissances de chaque pair est régénérée en se basant sur l'historique de toutes les requêtes émises.

### 3.2. Méthode de sélection des pairs

L'algorithme de sélection des pairs utilise la base $B(E_1,E_2)$, générée par le module de génération de profils pour choisir les pairs "pertinents". Pour une requête $Q$, l'algorithme détermine à partir de l'ensemble $E_1$, un ensemble de concepts similaires à $Q$ (noté par $SQTC$ : SimilarQueriesTermsConcepts), qui sera trié selon la valeur de la similarité. Par la suite, pour chaque concept $C_i$ de l'ensemble $SQTC$, il détermine, à partir de l'ensemble $E_2$, un ensemble de concepts similaires ($SQPC$ : SimilarQueriesPeersConcepts) contenant les meilleurs pairs pour la requête $Q$. Finalement, l'ensemble des pairs choisis est égal à l'ensemble des pairs figurants dans l'intension de chaque concept de l'ensemble $SQPC$. La complexité de l'algorithme $LPS$ présenté dans [1] est de $O(|E_2|\times|E_1|)$, avec $|E_i|$ dénote la cardinalité de l'ensemble $E_i$.

### 3.3. Limites

Bien que notre algorithme $LPS$ a donné des bons résultats par rapport aux autres algorithmes, il possède deux problèmes majeurs qui peuvent dégrader ces performances :

– La stratégie de mise à jour des bases de connaissances est une stratégie statiquequi consiste à regénérer la base de connaissances de chaque pair à partir de l'historique de toutes les requêtes émises. Cette stratégie est très coûteuse en terme de temps si le nombre d'objets (Requêtes) et d'attributs (termes dans le contexte $C1$ ou pairs dans dans le contexte $C2$) devient important ce qui dégrade les performances de l'algorithme de génération de concepts formels et pose des problèmes de passage à l'échelle de cette approche.

– La complexité de l'algorithme de sélection des pairs *LPS* est de $O(|E_2| \times |E_1|)$, avec $|E_i|$ dénote la cardinalité de l'ensemble $E_i$. Le problème majeur de cette solution est l'exploration importante d'un espace de recherche alors que la contrainte du seuil de nombre maximum des pairs à sélectionner n'a pas été exploitée.

## 4. Amélioration de l'algorithme *LPS*

### 4.1. Stratégie incrémentale de mise à jour de la base de connaissances

L'idée sous-jacente à notre proposition consiste à trouver une nouvelle stratégie de mise à jour de la base de connaissances permettant de prendre en compte les informations sur nouvelles requêtes sans dégrader les performances de l'algorithme de génération des concepts formels. Pour contourner ce problème, nous proposons de définir une stratégie incrémentale qui génère une base $B_+$ à partir de l'historique des nouvelles requêtes (les requêtes émises après la dernière opération de mise à jour). Par la suite, nous construisons la nouvelle base de connaissances en faisant l'union de $B_+$ et l'ancienne base. Avec cette stratégie, le nombre d'objets et d'attributs nécessaires pour la construction de la base $B_+$ est largement inférieur à celui de la stratégie statique. Par conséquent, le temps de mise à jour sera inférieur que celui de la stratégie statique.

### 4.2. Nouvelle version de l'algorithme *LPS*

Pour réduire l'espace de recherche, nous proposons de considérer la contrainte du seuil de nombre maximum des pairs à sélectionner $P_{max}$ dans le processus de construction des pairs sélectionnés en parallèle avec la sélection des concepts similaires à $Q$ (voir Algorithme1). Pour une requête $Q$, l'algorithme détermine à partir de l'ensemble $E_1$ le concept le plus similaire à $Q$, noté *concept* (fonction *getConcept*($B.E1,Q$) de l'algorithme 1). Par la suite, nous déterminons, à partir de l'ensemble $E_2$, un ensemble de concepts similaires à *concept*, noté *SQPC* (SimilarQueriesPeersConcepts) contenant les meilleurs pairs pourla requête$Q$ (fonction*getSimilarConcept*($B.E2,Extent(concept)$) de l'algorithme 1). L'ensemble des pairs choisis est égal à l'ensemble des pairs figurants dans l'intensionde chaqueconceptde l'ensemble*SQPC* (fonction*getSelectedPeers*($SQPC$) de l'algorithme 1. Si le nombre des pairs sélectionnés est inferieur à $P_{max}$ on sélectionne un autre concept, à partir de l'ensemble $E_1$, et on répète le même procédé. La complexité de la nouvelle version de notre algorithme est de $O(P_{max} \times (|E_1| + |E_2|))$ qui est inférieure à celle de l'algorithme *LPS*.

## 5. Intégration et expérimentations

### 5.1. Environnement

Pour tester notre approche, nous avons choisi le simulateur PeerSim [8], qui est un outil Opensource écrit en Java , présente l'avantage d'être déjà spécialisé pour l'étude des

systèmes P2P et dispose d'une architecture ouverte et modulaire qui permet de l'adapter
et de lui intégrer de nouvelles couches.

```
1  Algorithme : LEARNINGPEERSELECTION(B, Q, P_max)
2  Entrées :
3      B : Base de connaissances ;
4      Q : Requête ;
5      P_max : Nombre maximum des pairs à sélectionner ;
6  Sortie :
7      selectedPeers : liste des pairs sélectionnés
8  début
9      selectedPeers = ∅
10     concept = getConcept(B.E1, Q)
11     while concept et |selectedPeers| < P_max do
12         SQPC := getSimilarConcept(B.E2, Extent(concept))
13         selectedPeers := selectedPeers ∪ getSelectedPeers(SQPC)
14         B.E1 := B.E1 − {concept}
15         concept := getConcept(B.E1, Q)
16     Return (selectedPeers)
17 fin
```

**Algorithme 1** : NOUVELLE VERSION DE L'ALGORITHME DE SÉLECTION DE PAIRS

### 5.2. Les données source

Pour tester notre algorithme *LPS*, nous avons utilisé le jeu de données " BigDataSet ", développé dans le cadre du projet RARE. Ce jeu de données a été obtenu à partir d'une analyse statistique sur des données collectées d'un système pair-à-pair Gnutella [7] et des données de la collection TREC [13], ce qui nous permet de réaliser des simulations en conditions réelles.

"BigDataSet" est composé de 25000 documents et 4999 requêtes. Ils sont réparties et dupliquersur 499pairs. Ce jeu de donnéesest accompagnépar des fichiers XML décrivant les noeuds du système et les documents qu'ils possèdent, ainsi que les requêtes qui seront lancées sur le réseau.

### 5.3. Mesures d'évaluation

Pour tester les performances de notre approche, nous avons utilisé les métriques rappel (R), le trafic réseau en nombre de messages, le temps de génération des bases des connaissances et la cardinalité la base de connaissances en nombre de concepts. Le rappel (R) est définie comme suit pour une requête *Q*.

– $R(Q) = \frac{DPR}{DP}$ Avec,

– *DPR* : le nombre de documents pertinents retournés.

– *DP* : le nombre de documents pertinents.

## 5.4. Initialisation des paramètres de simulation

La simulation de notre algorithme ainsi que celui utilisé par Gnutella s'est basée sur les paramètres suivants :

– *TTL* : Profondeur maximale de recherche. Nous avons varié ce paramètre de 2 à 5 pour voir son impact sur la qualité de routage.

– *Pmax* : Nombre de pairs auxquels la requête doit être propagée, initialisé à 3.

– *Overlay size* : Nombre de pairs dans le réseau, initialisé à 500 (nombre de pairs dans le jeu de données).

De plus, notre algorithme a besoin d'une base de connaissances pour chaque pair. Pour cela, nous avons lancé les 20600 premières requêtes, en utilisant la méthode proposée dans Gnutella, afin de construire un fichier log initial pour chaque pair. Par la suite, nous avons lancé l'exécution du module de gestion de profils pour construire une base de connaissances initiale à partir de fichier log de chaque pair, noté $B_0$. Pour prendre en compte les nouvelles informations sur les requêtes passées, les bases des connaissances sont mises à jour après l'émission d'un nombre de requêtes.

## 5.5. Variation de *TTL* et impact sur les performances de *LPS* et de Gnutella

Pour comparer les performances de notre algorithme *LPS* par rapport à celui de Gnutulla, nous avons calculé le rappel moyen et la moyenne du nombre de messages par intervalle de 2000 requêtes envoyées par les différents pairs du système en variant le *TTL*. La figure 1 montre que le rappel moyen pour notre algorithme *LPS* est plus élevé que celui de Gnutella pour toutes les variations du *TTL*. En effet, le rappel moyen pour notre algorithme varie entre :

– 0,28 et 0,25 pour un *TTL* égal à 2
– 0,45 et 0,49 pour un *TTL* égal à 3
– 0,63 et 0,69 pour un *TTL* égal à 4
– 0,82 et 0,86 pour un *TTL* égal à 5

Tandis que le rappel moyen pour Gnutella varie entre 0,43 et 0,44 dans les meilleurs des cas pour un *TTL* égal à 5. De plus, le rappel de l'algorithme *LPS* augmente à chaque enrichissement des bases de connaissances.

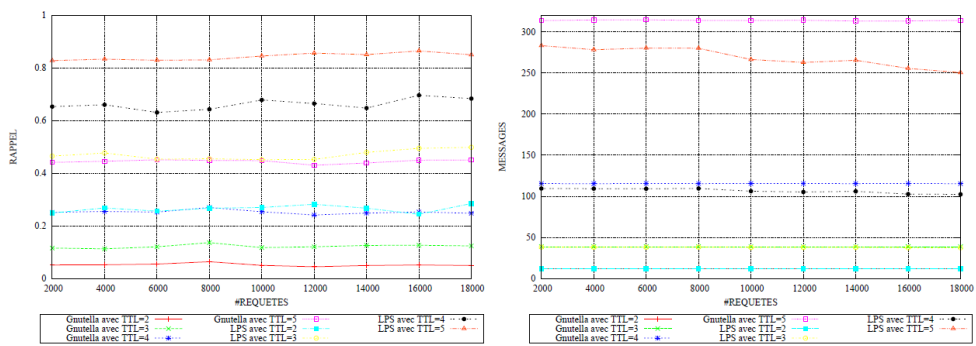

**Figure 1.** *Rappel en fonction du nombre de requêtes*  **Figure 2.** *Nombre de messages en fonction du nombre de requêtes*

La figure 2 montre que le nombre de messages pour l'algorithme *LPS* est inférieur à celui de Gnutella pour toutes les variations de *TTL*. Nous remarquons que la différence en nombre de messages est très faible pour un *TTL* inférieur à 3. Les deux algorithmes produisent presque le même nombre de messages, mais cette différence devient importante pour un *TTL* supérieur ou égal à 4. En effet, la moyenne du nombre de messages pour notre algorithme a passé de 283 en utilisant la base initiale $B_0$ à 262 en utilisant la base $B_0$ et à 259 en utilisant $B_2$ pour une valeur de *TTL* égale à 5. De même, ce nombre a passé de 109 en utilisant la base initiale $B_0$ à 106 en utilisant la base $B_1$ et à 102 en utilisant $B_2$ pour une valeur de *TTL* égale à 4. Par contre il reste stable pour Gnutella avec une valeur égale à 313 si le *TTL* est égal à 5 et à 115 si le *TTL* est égal à 4.

A partir de ces deux figures, nous pouvons déduire que la taille de la base de connaissances a un impact sur la qualité de routage de notre algorithme et que la qualité de routage de notre algorithme devient acceptable à partir d'une valeur de *TTL* supérieure ou égale à 3. Ainsi, nous remarquons qu'à partir de *TTL* égal à 3, le rappel moyen de notre algorithme *LPS* devient supérieur que celui de Gnutella avec un *TTL* égal à 5. Par conséquent, nous pouvons dire que l'algorithme *LPS* produit des bons résultats avec un nombre de messages moyen égal à 37. Alors que, Gnutella commence à produire des résultats acceptables avec un nombre moyen de messages égal à 313.

## 5.6. Comparaison des performances des stratégies de maintenance statique et incrémentale

Durant la simulation, les bases des connaissances sont mises à jour 2 fois. En effet, nous avons lancé les 20600 premières requêtes, en utilisant la méthode de Gnutella, afin de construire au niveau de chaque pair une base initiale, noté $B_0$. Par la suite, les bases des connaissances sont mises à jour après l'émission de 8000 requêtes pour construire la base $B_1$ puis après 6000 pour construire $B_2$. Pour comparer les performances de la méthode statique et la méthode incrémentale nous avons calculé le temps moyen nécessaire pour mettre à jour une base d'un pair. De plus, le nombre de concepts moyen de chaque base a été calculé.

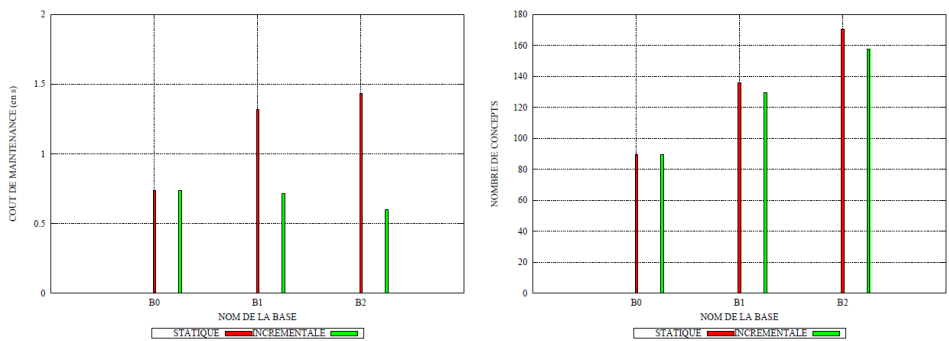

**Figure 3.** *Coût de maintenance en fonction de la taille de la base* **Figure 4.** *Nombre de concepts en fonction de la taille de la base*

La figure 3 montre que le temps moyen de mise à jour d'une base d'un pair avec la stratégie incrémentale est très inférieur que celui avec la stratégie statique. En effet, le coût de maintenance avec la stratégie statique augmente à chaque opération de mise à jour puisque le nombre de requêtes, utilisées pour mettre à jour la base, est égal à celui de l'ancienne base plus le nombre des nouvelles requêtes. Par contre, le coût de maintenance avec la stratégie statique ne dépend que du nombre des nouvelles requêtes. La figure 4 montre que le nombre de concepts générés par la stratégie incrémentale est toujours inférieur à celui générés par la stratégie statique. Le nombre de concepts joue un rôle très important dans l'algorithme *LPS*.

### 5.7. Impact de la stratégie de maintenance sur la qualité de routage

Pour étudier l'impact de deux stratégies de mise à jour de la base de connaissances d'un pair sur la qualité de routage de notre algorithme nous avons calculé le rappel moyen et la moyenne du nombre des messages par intervalle de 2000 requêtes envoyées par les différents pairs du système pour des *TTL* égaux à 3 et 5.

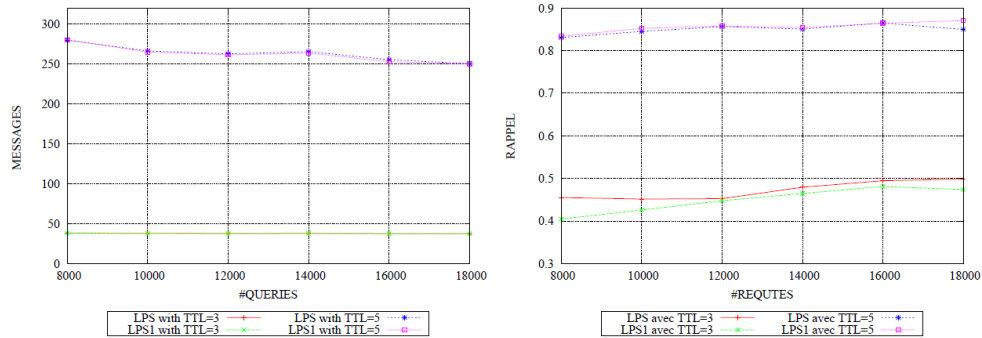

**Figure 5.** *Variation du nombre de messages* **Figure 6.** *Rappel en fonction du nombre de en fonction du nombre de requêtes selon les requêtes selon les deux stratégies deux stratégies*

La figure 5 montre que les deux stratégies donnent les mêmes performances en termes de rappel. *LPS* respectivement *LPS*1, dénote les résultats de notre algorithme en utilisant la stratégie statique, respectivement, la stratégie incrémentale. En effet, il y a une légère différence due à la sélection aléatoire de quelques pairs. Ainsi, la figure 6 montre que les deux stratégies donnent aussi le même nombre de messages.

## 6. Conclusion

De nombreux travaux de recherche proposent d'exploiter la sémantique sur le contenu des pairs pour router plus efficacement les requêtes, mais peu de travaux explorent les informations relatives à l'historique des requêtes. Ainsi, nous avons proposé une amélioration de l'algorithme de routage des requêtes par apprentissage *LPS*. Nous avons développé également, une nouvelle méthode incrémentale de mise à jour des bases de connaissances. Des tests, réalisés en utilisant le simulateur PeerSim, ont montré que notre algorithme de routage est plus performant qu'un algorithme de routage classique, en termes de rappel et de nombre de messages, et que la qualité de routage de notre algorithme dépend de la taille de la base de connaissances. De même nous pouvons dire, que la stratégie de maintenance incrémentale est plus efficace que la stratégie statique. Pour le prolongement de ce travail, nous proposons d'utiliser d'autres bases de tests pour mieux évaluer les performances de l'approche proposée.

## 7. Bibliographie